\documentclass[12pt]{article}
\begin{document}
\global\parskip 6pt
\newcommand{\be}{\begin{equation}}
\newcommand{\ee}{\end{equation}}
\newcommand{\bea}{\begin{eqnarray}}
\newcommand{\eea}{\end{eqnarray}}
\newcommand{\non}{\nonumber}

\begin{titlepage}
\hfill{hep-th/0205246}
\vspace*{1cm}
\begin{center}
{\Large\bf Bubbles in Anti-de Sitter Space}\\
\vspace*{2cm} Danny Birmingham\footnote{E-mail: danny.birmingham@ucd.ie}
and Massimiliano Rinaldi\footnote{Email: massimiliano.rinaldi@ucd.ie}\\
\vspace*{.5cm}
{\em Department of Mathematical Physics\\
University College Dublin\\
Belfield, Dublin 4, Ireland}\\
\vspace{2cm}

\begin{abstract}
We explore the bubble spacetimes which can be obtained from double
analytic continuations of static and rotating black holes in
anti-de Sitter space. In particular, we find that rotating black
holes with elliptic horizon lead to bubble spacetimes only in
dimension greater than five. For dimension greater than seven, the
topology of the bubble can be non-spherical. However, a bubble
spacetime is shown to arise from a rotating de Sitter black hole
in four dimensions. In all cases,
the evolution of the bubble is of de Sitter type. Double analytic
continuations of hyperbolic black holes and branes are also
discussed.
\end{abstract}
\vspace{1cm}
May 2002
\end{center}
\end{titlepage}
\vspace{1.5 cm}

\section {Introduction}
The formulation of string theory in time-dependent backgrounds
presents a particularly challenging problem, although progress can
be achieved by considering some simple time-dependent solutions.
As a step in this direction, a class of time-dependent backgrounds
has been investigated recently in \cite{Aharony}. The spacetimes
considered are obtained from a double analytic continuation of
asymptotically flat black holes, and describe the Lorentzian
evolution of a bubble. The technique of double analytic
continuation was originally developed for the study of the
stability of the Kaluza-Klein vacuum \cite{Witten}, see also
\cite{Dowker1,Dowker2}. This technique has also been used in the
formulation of a positive energy theorem for anti-de Sitter space
\cite{Horowitz}, and discussed within the context of brane world
scenarios \cite{Ida1,Ida2}, and M-theory \cite{Costa,Fabinger}.

In general, the evolution pattern of the bubble is determined by
the form of the original black hole metric. The analysis of
\cite{Aharony} focussed on a class of bubbles which arise from
asymptotically flat Kerr black holes. Indeed, the presence of the
rotation parameters is crucial in order to obtain interesting time
dependent behavior. It was found that at early times the bubble
undergoes a de Sitter phase of exponential expansion, while at
late times the evolution is governed by a milder Milne phase.

It is of interest to explore the possible time-dependent
spacetimes which can arise from double analytic continuation of
more general black holes. Our aim here is to investigate the
bubble spacetimes which can be obtained from the analytic
continuation of black holes in de Sitter and anti-de Sitter space.
In the latter case, there is wide class of topological black hole
spacetimes available, due to the possibility of non-trivial
horizon topology \cite{Lemos}-\cite{Birmingham}.
In particular, the topology of the horizon can be elliptic (which
includes the standard spherical case), toroidal, or hyperbolic.
The rotating counterparts of these solutions have also been
constructed \cite{Hawking}-\cite{Dehghani1}.
We find that the static and rotating black holes with elliptic
horizon lead to bubbles which have only a de Sitter phase of
evolution. Furthermore, the rotating black holes lead to
acceptable bubble spacetimes only in dimension $d\geq 6$. Moreover,
it is
possible to have a bubble with non-spherical (elliptic)
topology in dimension $d \geq 8$. For the case of hyperbolic
horizon, one can indeed perform the double analytic continuation.
However, in all cases, one finds that the evolution of the bubble
is determined by the embedding of an anti-de Sitter space, and
thus does not lead to an evolving bubble situation. Finally, we
show that a bubble spacetime  does arise from the rotating de
Sitter black hole in four dimensions.

\section{Static Anti-de Sitter Black Holes}
It is useful to begin with the static black hole in $d$ dimensions, with line
element given by
\bea
ds^{2} = -f(r) dt^{2} + f^{-1}(r) dr^{2} +
r^2 h_{ij}(x)dx^{i}dx^{j}, \label{static}
\eea
with
\bea
f = k + \frac{r^2}{l^2}- \frac{2M}{r^{d-3}}.
\eea
The topology of the
black hole horizon is labelled by the parameter $k$, which can be
normalized to take the values $\pm 1, 0$. The above metric is a
solution of Einstein's equations with negative cosmological
constant, $R_{\mu\nu} = -(d-1)/l^2 g_{\mu\nu}$. We consider the
class of locally asymptotically anti-de Sitter black holes, for
which the topology of the horizon is either elliptic ($k=1$),
toroidal ($k=0$), or hyperbolic ($k=-1$). The possibility of
anti-de Sitter black hole solutions with non-trivial topology
$k=0,-1$ was first discussed in four dimensions
\cite{Lemos}-\cite{Brill}, and generalized to
arbitrary dimensions in \cite{Birmingham}.

For the case of $k=1$, the line element of the horizon space is
given by $h_{ij}(x)dx^{i}dx^{j} = d\Omega_{d-2}^2$. For
definiteness, we consider the spherical case where
$d\Omega_{d-2}^2 = d\theta^2 + \sin^{2}\theta \;d\Omega_{d-3}^2$.
The associated bubble spacetime is obtained in the standard manner
\cite{Witten} by performing the double analytic continuation $t =
i\chi, \theta = \frac{\pi}{2} + i\tau$. This leads to the bubble
spacetime
\bea
ds^2 = f d\chi^2 + f^{-1}dr^2 +
r^2\left(-d\tau^{2}+\cosh ^2\tau\; d\Omega^2_{d-3}\right).
\eea
The radial variable is now restricted to the range $r \geq r_{+}$,
where $r_{+}$ is the largest real zero of $f$. Regularity of the
metric at $r_{+}$ then requires that $\chi$ be identified as a
periodic variable with period given by \cite{Birmingham}
\bea
\beta=\frac{4\pi l^2r_{+}}{(d-1)r_{+}^{2} + (d-3)kl^2}.
\eea
In
the usual way \cite{Witten}, this spacetime now describes a
bubble at $r = r_{+}$ evolving in an asymptotically anti-de Sitter
spacetime. We see that the bubble grows exponentially in time and
the geometry traced out by the $r = r_{+}$ surface is a
$(d-2)$-dimensional de Sitter spacetime.

The horizon metric for the $k=-1$ case can be written in the form
$h_{ij}(x) dx^{i}dx^{j} = d\Sigma_{d-2}^2 = d\theta^{2} + \sinh^{2}\theta
\;d\Omega_{d-3}^{2}$. If we continue $\theta=i\tau$, we obtain a
metric that clearly has non-Lorentzian signature. However, we can
obtain a Lorentzian bubble spacetime if we continue an angular
variable in $d\Omega_{d-3}^{2} = (d\psi^{2} + \sin^{2}\psi\;
d\Omega_{d-4}^{2})$. Upon the substitution $\psi=i\tau+\pi/2$, the
metric becomes
\bea
ds^2=fd\chi^2+f^{-1}dr^2+r^2\left[d\theta^2+\sinh ^2\theta
\left(-d\tau^2+\cosh ^2\tau \;d\Omega^2_{d-4}\right)\right].
\eea
In this case, one notes that the term in square brackets describes
a $(d-2)$-dimensional anti-de Sitter space, which can be written in globally
static coordinates. Thus, the bubble
spacetime in this case is static. This case has also been considered in the
context of a positive energy theorem in \cite{Horowitz}.

\section{Rotating Anti-de Sitter Black Holes}
In \cite{Aharony}, a more interesting class of time-dependent
backgrounds was obtained through the analytic continuation of
rotating black holes in four and higher dimensions. In particular,
the presence of the rotating parameters allowed an easing of the
exponential expansion of the bubble into a milder Milne phase. In
order to check whether such behavior is also present in the
anti-de Sitter case, we begin with the line element of the
rotating black hole with elliptic horizon. This takes the form
\cite{Hawking}-\cite{Dehghani1}
\bea
\label{rotell}
ds^2 &=&
-\frac{\Delta_r}{\rho^2}\left[dt-\frac{a}{\Xi}\sin^2\theta
d\phi\right]^2+\frac{\rho^2}{\Delta_{r}}dr^2+\frac{\rho^2}
{\Delta_{\theta}}d\theta^2 +\frac{\Delta_{\theta}}{\rho^2}\sin
^2\theta\left[a dt-\frac{r^2+a^2}{\Xi}d\phi\right]^2 \non\\
&+&r^2\cos ^2\theta\; d\Omega_{d-4}^2,
\eea
where $a$ is the
angular momentum. Again, for definiteness, we shall consider the
case of spherical horizon, for which $d\Omega_{d-4}^2$ describes
the metric of a $(d-4)$-sphere. Of course, this
$(d-4)$-dimensional part of the spacetime can be replaced by a
general elliptic space, for example a lens space. In the above, we
have
\bea
\Delta_r &=&
(r^2+a^2)\left(1+\frac{r^2}{l^2}\right)-\frac{2M}{r^{d-5}},\non\\
\Delta_{\theta} &=& 1-\frac{a^2}{l^2}\cos ^2\theta, \non \\
\Xi &=&1-\frac{a^2}{l^2}, \non\\
\rho^2 &=& r^2+a^2\cos ^2\theta.
\eea

The standard Euclidean
section of the black hole is defined by the analytic continuation
$t=i\chi$ and $a = -i \alpha$. The resulting spacetime is
defined for $r \geq r_{+}$, where $r_{+}$ is now the largest real
zero of $\Delta_{r}$. In the usual way, regularity of the metric
at $r_{+}$ then requires the coordinate identifications
\bea
(\chi,\phi) \equiv (\chi + 2 \pi R n_{1}, \phi + 2 \pi R\Omega n_{1} + 2 \pi n_{2}),
\eea
where $n_{1}, n_{2} \in {\bf Z}$. We have
\bea
R =  \frac{2(r_+^2-\alpha^2)}{\Delta^{\prime}_r(r_+)},\;\;
\Omega=\frac{\alpha\;\Xi}{(r^2_+ -\alpha^2)},
\eea
where $R$ is the inverse of the surface gravity \cite{Hawking} after the continuation.
Let us now apply the
analytic  continuation $\theta=i\tau+\pi/2$ in four dimensions.
The main observation here is that the metric has a time-dependent
signature. Indeed, the induced metric on the bubble, at $r=r_+$,
is given by
\bea
ds^2_{bubble}&=&-\left[\frac{r_+^2+\alpha^2\sinh^2\tau}{1-\frac{\alpha^2}{l^2}\sinh
^2\tau}\right]d\tau^2 + \left[\frac{1-\frac{\alpha^2}{l^2}\sinh
^2\tau}{r_+^2+\alpha^2\sinh ^2\tau}\right]\left(\frac{\alpha}{\Omega}\right)^{2}
\cosh ^2\tau \;d\tilde \phi^2,
\eea
where $\tilde{\phi} = \phi - \Omega \chi$.
One then sees that there exists a value $\tau_{crit}$ such that
the induced metric has signature $(-,+)$ for $|\tau|<\tau_{crit}$,
and $(+,-)$ otherwise. Thus, the four-dimensional case does not
yield a suitable bubble spacetime. Clearly, the influence of the
anti-de Sitter radius $l^{2}$ is crucial to this behavior, and the
metric reduces to the one found in \cite{Aharony} in the limit
$l^{2} \rightarrow \infty$.

Proceeding to higher dimensions, one
notes that an acceptable double analytic
continuation is possible by choosing to continue one of the coordinates
in the $d \Omega_{d-4}^2$ part of the metric.
This leads to the bubble spacetime
\bea
ds^2&=&\frac{\Delta_r}{\rho^2}\left[d\chi+ \frac{\alpha}{\Xi}\sin^2\theta
d\phi\right]^2+\frac{\rho^2}{\Delta_{r}}dr^2+\frac{\rho^2}
{\Delta_{\theta}}d\theta^2 +\frac{\Delta_{\theta}}{\rho^2}\sin
^2\theta\left[\alpha
d\chi - \frac{(r^2-\alpha^2)}{\Xi}d\phi\right]^2+ \non \\
&+& r^2\cos ^2\theta \left[-d\tau^2+\cosh ^2\tau\;
d\Omega_{d-5}^2\right],
\eea
where
\bea
\Delta_r &=&
(r^2-\alpha^2)\left(1+\frac{r^2}{l^2}\right)-\frac{2M}{r^{d-5}}, \non \\
\Delta_{\theta} &=&1+\frac{\alpha^2}{l^2}\cos ^2\theta,
\non \\
\Xi &=&1+\frac{\alpha^2}{l^2}, \non \\
\rho^2&=&r^2-\alpha^2\cos
^2\theta.
\eea
The time-dependent part of the bubble geometry is now
described by the embedding
of a $(d-4)$-dimensional de Sitter space.
Therefore, the bubble spacetime is time-dependent
only if $d \geq 6$. Clearly, the $(\theta,\phi)$ part of the
spacetime does not take part in this evolution.
Thus, in contrast to the models obtained in \cite{Aharony},
the anti-de Sitter
case does not yield phases of milder evolution.
Furthermore, one sees that for $d\geq 8$, the $d\Omega_{d-5}^{2}$ part of the
bubble metric can be a general elliptic space.

The analogous rotating solution with hyperbolic topology
is given by \cite{Klemm}
\bea
ds^2&=&-\frac{\Delta_r}{\rho^2}\left[dt+\frac{a}{\Xi}\sinh^2\theta
d\phi\right]^2+\frac{\rho^2}{\Delta_{r}}dr^2+\frac{\rho^2}
{\Delta_{\theta}}d\theta^2 +
\frac{\Delta_{\theta}}{\rho^2}\sinh ^2\theta\left[a
dt-\frac{(r^2+a^2)}{\Xi}d\phi\right]^2\non\\
&+&r^2\cosh ^2\theta\; d\Sigma_{d-4}^2,
\eea
where
\bea
\Delta_r &=&
(r^2+a^2)\left(-1+\frac{r^2}{l^2}\right)-\frac{2M}{r^{d-5}}, \non \\
\Delta_{\theta} &=&1+\frac{a^2}{l^2}\cosh ^2\theta, \non \\
\Xi &=&1+\frac{a^2}{l^2}, \non \\
\rho^2&=&r^2+a^2\cosh ^2\theta,
\eea
and
$d\Sigma_{d-4}^2=d\psi^2+\sinh ^2\psi\; d\Omega_{d-5}^2$ is the
metric on a $(d-4)$-dimensional hyperbolic space. In this case,
the $(\theta,\phi)$ sector of the spacetime is necessarily non-compact \cite{Klemm}.
Replacing
$t=i\chi$ and $a = -i \alpha$, and imposing
the appropriate identifications on $\chi$
and $\phi$, a further continuation of $\theta$ again leads to a
metric with time-dependent signature for any $d\geq4$. However, we
can perform a continuation of an angular variable in the spherical
section embedded in $d\Sigma^2_{d-4}$,  and we find
\bea
ds^2&=&\frac{\Delta_r}{\rho^2}\left[d\chi-\frac{\alpha}{\Xi}\sinh^2\theta
d\phi\right]^2+\frac{\rho^2}{\Delta_{r}}dr^2+\frac{\rho^2}
{\Delta_{\theta}}d\theta^2 +\frac{\Delta_{\theta}}{\rho^2}\sinh
^2\theta\left[\alpha d\chi -\frac{(r^2-\alpha^2)}{\Xi}d\phi\right]^2+ \non \\
&+& r^2\cosh ^2\theta \left[d\psi^2+\sinh
^2\psi\left(-d\tau^2+\cosh ^2\tau\; d\Omega_{d-6}^2\right)\right],
\eea
with
\bea
\Delta_r &=&
(r^2-\alpha^2)\left(-1+\frac{r^2}{l^2}\right)-\frac{2M}{r^{d-5}}, \non \\
\Delta_{\theta} &=&1-\frac{\alpha^2}{l^2}\cosh ^2\theta,
\non \\
\Xi &=&1-\frac{\alpha^2}{l^2}, \non \\
\rho^2&=&r^2-\alpha^2\cosh^2\theta.
\eea
As in the static case with hyperbolic horizon,
the bubble spacetime is again described simply by the embedding of a
$(d-4)$-dimensional anti-de Sitter space.

One can also consider the $5$-dimensional anti-de Sitter black hole with
two rotational parameters \cite{Hawking}. In this case, one finds
that the only consistent continuation involves the angular
variable not related to rotations, which again leads to a
time-dependent signature. Finally, the analysis of the double
analytical continuation of a cylindrical rotating black hole
\cite{Klemm} is very similar to the $k=1$ case. In four
dimensions, the only possible continuation leads to a
time-dependent signature, while in higher dimensions one can
continue an angular variable of the spherical section.

\section{Rotating de Sitter Black Holes}

A rotating black hole in de Sitter space with one angular momentum
can be found by replacing $l^2\rightarrow -l^2$ in the metric
(\ref{rotell}), see for example \cite{Cai,Dehghani2}. In four
dimensions, we find behavior which contrasts the
anti-de Sitter case. Indeed, we find that the metric obtained
after the double analytic continuation has a constant signature.
To see this, we consider the metric (\ref{rotell}) for $d=4$,  and
make the continuations $l\rightarrow il$, $t=i\chi$, $a = -i\alpha$
and $\theta=i\tau+\pi/2$ (the choice $\theta=i\tau$ leads to a
time-dependent signature). The resulting metric takes the form
\bea
ds^2&=&\frac{\Delta_r}{\rho^2}\left[d\chi+ \frac{\alpha}{\Xi}\cosh^2\tau
d\phi\right]^2+\frac{\rho^2}{\Delta_{r}}dr^2-\frac{\rho^2}
{\Delta_{\tau}}d\tau^2
+\frac{\Delta_{\tau}}{\rho^2}\cosh^2\tau\left[\alpha
d\chi - \frac{(r^2-\alpha^2)}{\Xi}d\phi\right]^2,\;\;\non\\
\eea
where
\bea
\Delta_r &=& (r^2-\alpha^2)\left(1-\frac{r^2}{l^2}\right)-2Mr,\non \\
\Delta_{\tau} &=&1+\frac{\alpha^2}{l^2}\sinh^2\tau, \non \\
\Xi &=&1-\frac{\alpha^2}{l^2}, \non \\
\rho^2&=&r^2+\alpha^2\sinh ^2\tau.
\eea

The induced metric on the bubble for small $\tau$ is given by
\bea
ds^2 \simeq -r_+^2d\tau^2+\frac{\alpha^2}{\Omega^2 r_+^2}\cosh^2\tau
d\tilde\phi^2,
\eea
where $\tilde\phi =\phi - \Omega \chi$ with
$\Omega = \alpha \Xi/(r_{+}^2- \alpha^2)$. We see that the bubble
metric behaves like a 2-dimensional de Sitter space. For large
$\tau$, we have
\bea
ds^2\simeq
-l^2d\tau^2+\frac{\alpha^2}{\Omega^2 l^2}e^{2\tau} d\tilde\phi^2,
\eea
which
again looks like a de Sitter space at late times.
Thus, the evolution of the bubble is described by a de Sitter
phase only.

\section{Conclusion}
We have examined the double analytic continuations of static and
rotating black holes in anti-de Sitter and de Sitter spacetime.
The set of coordinates which can be continued is constrained by
the requirement that the resulting bubble spacetime be real
Lorentzian and time-dependent. The examples studied only led
to a de Sitter phase of time dependence, and do not appear to
support the milder phase of Milne evolution found in the
asymptotically flat case \cite{Aharony}. It would be worthwhile to
consider the classical and quantum stability of these spacetimes
along the lines of discussed in \cite{Aharony}.  \\

\noindent {\bf \large Acknowledgements}. M.R. was supported by Enterprise
Ireland grant BR/1999/031, and P.E. 2000/2002 from Bologna University.


\begin{thebibliography}{99}
\bibitem{Aharony} O. Aharony, M. Fabinger, G.T. Horowitz and E. Silverstein,
``Clean Time-dependent String Backgrounds from Bubble Baths,''
hep-th/0204158.
\bibitem{Witten} E. Witten, Nucl. Phys. B {\bf 195}, 481 (1982).
\bibitem{Dowker1} F. Dowker, J.P. Gauntlett, G.W. Gibbons and G.T. Horowitz,
Phys. Rev. D {\bf 52}, 6929 (1995); hep-th/9507143.
\bibitem{Dowker2} F. Dowker, J.P. Gauntlett, G.W. Gibbons and G.T. Horowitz,
Phys. Rev. D {\bf 53}, 7115 (1996); hep-th/9512154.
\bibitem{Horowitz} G.T. Horowitz and R.C. Myers, Phys. Rev. D {\bf 59},
026005 (1999); hep-th/9808079.
\bibitem{Ida1} D. Ida, T. Shiromizu and H. Ochiai,
Phys. Rev. D {\bf 65}, 023504 (2002); hep-th/0108056.
\bibitem{Ida2} H. Ochiai, D. Ida and T. Shiromizu, ``Quantum Creation of the
Randall-Sundrum Bubble,'' hep-th/0111070.
\bibitem{Costa}
M.S. Costa and M. Gutperle, JHEP {\bf 0103}, 027 (2001);
hep-th/0012072.
\bibitem{Fabinger}
M. Fabinger and P. Horava, Nucl.\ Phys.\ B {\bf 580}, 243 (2000);
hep-th/0002073.
\bibitem{Lemos}
J.P. Lemos, Phys.\ Lett.\ B {\bf 353}, 46 (1995); gr-qc/9404041.
\bibitem{Huang}
C.G. Huang and C.B. Liang, Phys.\ Lett.\ A {\bf 201}, 27 (1995).
\bibitem{Cai2}
R.G. Cai and Y.Z. Zhang, Phys.\ Rev.\ D {\bf 54}, 4891 (1996);
gr-qc/9609065.
\bibitem{Beng} S. {\AA}minneborg, I. Bengtsson, S. Holst and P.
Peld\'{a}n, Class. Quantum Grav. {\bf 13}, 2707 (1999);
gr-qc/9604005.
\bibitem{Mann} R.B. Mann, Class. Quantum Grav. {\bf 14}, L109 (1997);
gr-qc/9607071.
\bibitem{Vanzo} L. Vanzo, Phys. Rev. D {\bf 56}, 6475 (1997); gr-qc/9705004.
\bibitem{Brill} D.R. Brill, J. Louko and P. Peld\'{a}n,
Phys. Rev. D {\bf 56}, 3600 (1997); gr-qc/9705012.
\bibitem{Birmingham} D. Birmingham,
Class. Quantum Grav. {\bf 16}, 1197 (1999); hep-th/9808032.
\bibitem{Hawking}
S.W. Hawking, C.J. Hunter and M.M. Taylor-Robinson,
Phys. Rev. D {\bf 59}, 064005 (1999);
hep-th/9811056.
\bibitem{Klemm2}
D. Klemm, V. Moretti and L. Vanzo, Phys. Rev. D {\bf 57}, 6127 (1998);
Erratum - ibid. D{\bf 60}, 109902 (1999); gr-qc/9710123.
\bibitem{Klemm}
D. Klemm, JHEP {\bf 9811}, 019 (1998); hep-th/9811126.
\bibitem{Dehghani1}
M.H. Dehghani,``Rotating Topological Black Branes in Various
Dimensions and AdS/CFT  Correspondence,'' hep-th/0203049.
\bibitem{Cai}
R.G. Cai, Nucl. Phys. B {\bf 628}, 375 (2002); hep-th/0112253.
\bibitem{Dehghani2}
M.H. Dehghani, Phys. Rev. D {\bf 65}, 104030 (2002);
hep-th/0201128.
\end{thebibliography}
\end{document}